# Symmetry Breaking and Order in the Age of Quasicrystals

Ron Lifshitz[a]

*To Danny Shechtman, for enriching our lives so profoundly.*

**Abstract:** The discovery of quasicrystals has changed our view of some of the most basic notions related to the condensed state of matter. Before the age of quasicrystals, it was believed that crystals break the continuous translation and rotation symmetries of the liquid-phase into a discrete lattice of translations, and a finite group of rotations. Quasicrystals, on the other hand, possess no such symmetries—there are no translations, nor, in general, are there any rotations, leaving them invariant. Does this imply that no symmetry is left, or that the meaning of symmetry should be revised? We review this and other questions related to the liquid-to-crystal symmetry-breaking transition using the notion of indistinguishability. We characterize the order-parameter space, describe the different elementary excitations, phonons and phasons, and discuss the nature of dislocations—keeping in mind that we are now living in the age of quasicrystals.

**Keywords:** quasicrystals · indistinguishability · symmetry · phasons · dislocations · point groups

## 1. Introduction

The year 2012 marks the centennial of the discovery of the diffraction of x-rays by crystals. This discovery was a great triumph not only for the idea that matter is composed of atoms, but also for the underlying paradigm of modern crystallography, stating that all *structurally ordered matter* is composed of periodic arrangements of these atoms. Since its early days, modern crystallography treated *order* and *periodicity* synonymously, both serving equally to define the notion of a crystal. With that came the so-called "crystallographic restriction," stating that crystals cannot have certain forbidden symmetries, such as 5-fold rotations. The periodicity of crystals became the underlying paradigm not only for crystallography itself, but also more generally for materials science and solid-state physics or chemistry, whose most basic tools relied on periodicity.

The year 2012 also marks the 30[th] anniversary of the discovery of quasicrystals, which completely shattered this paradigm, leading to a rebirth of crystallography.[1] Cahn[2] described the discovery of quasicrystals as a Kuhnian scientific revolution,[3] and I believe that we are now in the midst of the most exciting stage of this revolution. The old established paradigms, most importantly that of the periodicity of crystals, have been overthrown. The initial skepticism of the scientific community—embodied most vividly in the writings of Pauling[4]—is now gone, replaced by full recognition, with the award of the 2011 Nobel Prize in Chemistry to Shechtman, which we are celebrating with this festive Issue. New notions and paradigms are being tested and carefully adopted. New theories of quasicrystals, aperiodic tilings, and symmetry are being developed. Experimental techniques are undergoing fundamental modifications to encompass aperiodic crystals. All this intense activity is being pursued by hundreds of scientists worldwide, ranging from pure mathematicians and crystallographers to physicists, chemists, materials scientists, and even a few architects.

I would like to share some of the exciting surprises encountered in the current paradigm-building phase we are in by reviewing the successful adaptation of a number of fundamental notions—related to symmetry breaking and order—to the age of quasicrystals. Order, or more specifically long-range order, is a well-established notion, used extensively in theories of phase transitions. The emergence of order is associated with a spontaneous breaking of symmetry, where the less-ordered phase is characterized by a higher degree of symmetry than the ordered one. In fact, it is the change in symmetry that distinguishes between the two phases. In the case of crystals we are talking about the positional order of the atoms. The isotropic pairwise interactions between atoms give rise to a free energy that is isotropic and translationally invariant—one can

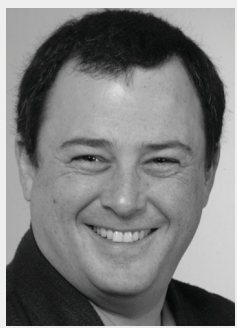

Born in Jerusalem, Ron Lifshitz received his Ph.D. in physics from Cornell University in 1995, where he studied the symmetry of quasiperiodic crystals with David Mermin. He then spent four years as a Research Fellow at Caltech, after which he assumed a faculty position at Tel Aviv University. His interests in quasicrystals range from fundamental questions such as order and symmetry, through physical questions of thermodynamic stability and electronic structure, to photonic applications based on metamaterials with quasicrystalline order.

[a] Ron Lifshitz
Raymond and Beverly Sackler School of Physics & Astronomy, Tel Aviv University
Tel Aviv 69978, Israel
e-mail: ronlif@tau.ac.il

shift or rotate space by any amount and the free energy remains unchanged. The disordered liquid phase has all these symmetries of the free energy, but the ordered crystal phase breaks this symmetry—it is no longer invariant under arbitrary translations or rotations.

What may seem surprising at first sight is the fact that a physical state—that of the ordered phase—has less symmetry than the physical interactions that give rise to it. This seems to contradict the general belief—articulated by Curie[5] in his famous symmetry principle—that "when effects show a certain asymmetry, this asymmetry must be found in the causes which give rise to them." Nevertheless, the idea is not new, and dates at least as far back as Euler's description of the buckling of a compressed elastic beam. Today, spontaneous symmetry breaking serves as one of the most basic notions of condensed-matter physics,[6] as well as being the underlying basis for the explanation of the so-called Higgs mechanism in particle physics.[7]

The general framework for treating the emergence of long-range order was formulated by Landau in his theory of phase transitions.[8] As described very clearly by Sethna,[9] it consists of providing answers to a number of basic questions: (1) *What is the appropriate order parameter with which one can measure the degree of order?* Such a quantity should be zero in the disordered phase and nonzero in the ordered phase. (2) *What is the broken symmetry?* To answer this, one often uses the language of group theory, characterizing the symmetry of the free energy, or the disordered phase, by a group *G*; the symmetry of the order parameter, or of the ordered phase, by a subgroup *H*; and then, if *H* is a normal subgroup, associating all the broken symmetries that are in *G* but not in *H* with the quotient group *G/H*. Applying the broken-symmetry operations to the physical state in the ordered phase changes it into a different, yet energetically equivalent state, as these operations do not change the free energy. Thus, the quotient group *G/H* is very useful in mapping out all the degenerate minimum free-energy states, making up the so-called "order-parameter space". (3) *What are the elementary excitations?* These are low-energy excitations, taking the free energy just above its minimum value. They are the so-called *Goldstone modes* that are directly related to the broken symmetry and to the nature of the order-parameter space. (4) *What are the topological defects?* These are imperfections in the otherwise perfectly ordered state, also directly related to the nature of the order-parameter space.

In what follows we shall answer these questions in the context of the liquid-to-crystal phase transition, where it is the *breaking of translational symmetry* that plays the important role,[9] keeping in mind that we are now living in the age of quasicrystals. The discussion will take us through a redefinition of the term *crystal* (Sec. 2); a reassessment of what we actually mean when we say that a crystal has a certain *symmetry* (Sec. 3); the introduction of a new type of elementary excitation—the *phason* (Sec. 4); and a generalization of the notion of a *dislocation* (Sec. 5), where one can no longer speak of the simple termination of one plane in a sequence of periodically ordered planes of atoms.

## 2. What is the order parameter for a crystal?

We wish to find an appropriate order parameter for crystals in light of the fact that we now know that ordered matter need not be periodic. To do so we must first understand what we mean when we say that atoms are ordered in space. Roughly speaking, the crystalline order parameter should be able to measure the extent of our ability to describe the positions of atoms in far away regions of space, based on our knowledge of their positions nearby. More technically, it should be a measure of the correlations between the positions of atoms in distant regions of the material. If $\rho(\mathbf{r})$ is a function, measuring the deviation of the density in a certain material from its average value, then the function that measures correlations between two points separated in space by a vector $\mathbf{R}$ —the two-point autocorrelation function—is known in crystallography as the *Patterson function*,[10]

$$P(\mathbf{R}) = \lim_{V \to \infty} \frac{1}{V} \int_V \rho(\mathbf{r}-\mathbf{R})\rho(\mathbf{r}) d\mathbf{r} . \qquad (1)$$

Luckily, this expression is directly related to the intensity, measured in diffraction experiments, where x-rays, electrons, neutrons, or any other quantum probes, are elastically scattered by the material. The measured intensity, $I(\mathbf{k}) = |\rho(\mathbf{k})|^2$, is simply the Fourier transform of (1), where $\rho(\mathbf{k})$ is the Fourier coefficient of the density $\rho(\mathbf{r})$ at wave vector $\mathbf{k}$. This can provide some physical intuition regarding the notion of long-range positional order, beyond that which is obtained through periodicity. It also gives us a particularly useful tool for characterizing ordered matter.

When scattered waves reach the detector—a photographic plate or a CCD camera—at a given position, they interfere with each other. If the atoms are arranged randomly in space, the relative phases of the detected waves will be random, and the waves will interfere destructively, producing not more than a very weak signal everywhere. On the other hand, if the atoms are ordered, arranged in correlated positions, one could imagine that at special and precise points on the detector they could interfere constructively, giving rise to intense peaks. If the intensity of these peaks scales as the number *N* of atomic scatterers, we can conclude that the positions of all the *N* atoms must be carefully arranged and correlated such that all *N* waves arrive at the detector with the same phase. This is the defining property of the familiar *Bragg peaks*, from which one extracts the information needed for crystal-structure determination. In the thermodynamic limit ($N \to \infty$) these peaks will diverge like Dirac delta functions.

Indeed, it was realized long ago[11] that a very convenient order parameter that signals a transition from a disordered liquid to an ordered solid—indicating the emergence of non-trivial correlations—appears in the form of delta functions, or Bragg peaks, at non-zero wave vectors in the diffraction diagram. We shall use this order parameter. In the case of a periodic crystal these wave vectors form a periodic *lattice,* reciprocal to the lattice of translations in real space leaving the crystal invariant. It was a great surprise, three decades ago, to realize that the condition of constructive interference from all the atoms can be fulfilled even when the atoms are not arranged periodically, as in the Penrose tiling[12] of Fig. 1. This was observed, of course, in Shechtman's experiment,[13] as well as in the theoretical work of Mackay[14] and later also of Levine and Steinhardt,[15] Elser,[16] and others. Elser even showed very early on that a particular kind of disorder, associated with the phason degrees of freedom, discussed below, does not destroy the Bragg peaks, even though it adds diffuse scattering.

There is an issue of semantics to consider before moving on. Before the age of quasicrystals, both periodicity and order were used to define the term *crystal*. After the discovery, one had to decide on a new definition. Although there is still a bit of inconsistency in the literature, the scientific community tends to associate the term crystal with having long-range order. Thus, one speaks of *periodic crystals* and of *aperiodic crystals*. The first official step in this direction was made by the International Union of Crystallography, through its Commission on Aperiodic Crystals,[17] which in its report for the year 1991 introduced an empirical definition of the term crystal that abolishes periodicity, and shifts the focus from a microscopic description of the actual crystal to its measured diffraction diagram. Realizing that Bragg peaks are observed in all known crystals, the Commission stated that a crystal is *"any solid having an essentially discrete diffraction diagram."* More recent discussions[18,19] have made the definition more explicit by stating that *"a crystal is a solid that has long-range positional order,"* and then considering what it means to have such order. The reader may want to refer to Ref. [19] for a discussion on the consequences of associating order with spatial correlations. It turns out that the real-space characterization of sets of points that produce Bragg peak diffraction has proven to be quite elusive, yielding many surprising and unintuitive results along the way. It is still a very active field of research.[20]

## 3. What is the broken symmetry?

### 3.1. Symmetry of the disordered phase – a liquid

As we are considering the transition from a liquid to an ordered solid, the symmetry of the disordered phase, and therefore presumably of the free energy as well, is given by the Euclidean group. This is the group of all rigid translations and rotations of space.[†] But, we emphasize once again that we shall focus on the *broken translational symmetry*,[9] while considering the breaking of rotational symmetry as a direct consequence of the former.

The basic idea, at least for continuous phase transitions, is that, at the transition, the deviations $\rho(\mathbf{r})$ of the density from its average value are very small. This justifies expanding the free energy $F$ in powers of $\rho(\mathbf{r})$ and its derivatives, as long as all terms in the expansion are invariant under the Euclidean group. Thus, for example, the term $\nabla\rho$ is not allowed, while $(\nabla\rho)^2$ and $\nabla^2\rho$ are fine. As the free energy contains entropic contributions, some of its expansion coefficients are expected to depend on temperature. Above $T_c$, in the liquid phase, the minimum value of the free energy is zero, and is obtained for $\rho(\mathbf{r}) = 0$ everywhere. As one cools down the material, and crosses below $T_c$, the coefficients of the different terms in the free energy vary, allowing the free energy of a state with a non-zero and non-uniform $\rho(\mathbf{r})$ to become negative, thus effecting a symmetry-breaking transition.

### 3.2. Symmetry of the ordered phase – a quasiperiodic crystal

#### 3.2.1. What is a quasiperiodic crystal?

We said that the ordered solid is characterized by having Bragg peaks in its experimental diffraction diagram, or equivalently in its calculated Fourier spectrum. We now wish to be more specific and limit ourselves to a particular subcategory of structures satisfying this requirement, namely to quasiperiodic crystals. For specific terminology we follow H. Bohr's theory of almost periodic functions,[21] and say that an *almost periodic crystal* is a solid whose density function $\rho(\mathbf{r})$ may be expressed as a superposition of a *countable* number of plane waves

$$\rho(\mathbf{r}) = \sum_{\mathbf{k} \in L} \rho(\mathbf{k}) e^{i\mathbf{k}\cdot\mathbf{r}} . \qquad (2)$$

In particular, if taking integral linear combinations of a finite number $D$ of wave vectors in this expansion can span all the rest, then the crystal is *quasiperiodic*. Owing to its finite resolution, an experimental diffraction pattern of a real quasiperiodic crystal will exhibit Bragg peaks only on a finite subset of $L$, each of which can be indexed by $D$ integers. If $D$ is the smallest number of wave vectors that can span the whole set $L$ using integral linear combinations, then $D$ is called the *rank*, or the *indexing dimension* of the crystal. *Periodic crystals* form a special subset of all quasiperiodic crystals, whose rank $D$ is equal to the physical dimension $d$ (the number of components in the vectors $\mathbf{r}$ and $\mathbf{k}$). The term *quasicrystal*, first introduced by Levine and Steinhardt,[15] is short for quasiperiodic crystal, but is commonly used to refer to those quasiperiodic crystals that are strictly aperiodic, with $D > d$.[‡]

#### 3.2.2. The notion of indistinguishability

We limit our discussion to ordered phases that are quasiperiodic, and continue by following Mermin's line of argument.[24] Thus, we consider crystals whose expansion is given by Eq. (2) with a finite rank $D$. These contain periodic crystals ($D = d$) and quasicrystals ($D > d$) as two distinct subsets. By substituting the sum of density waves (2) into an expansion of the free energy in powers of $\rho(\mathbf{r})$ and its derivatives, we obtain an alternative expression for the free energy in Fourier space,

$$F = \sum_n \sum_{\mathbf{k}_1,\ldots,\mathbf{k}_n \in L} A(\mathbf{k}_1,\ldots,\mathbf{k}_n)\rho(\mathbf{k}_1)\cdots\rho(\mathbf{k}_n), \qquad (3)$$

where the wave-vector dependence of the coefficients $A(\mathbf{k}_1,\ldots,\mathbf{k}_n)$ originates from the gradient terms in the original expansion. These coefficients must vanish, unless $\mathbf{k}_1 + \ldots + \mathbf{k}_n = 0$, as an immediate consequence of the invariance of $F$ with respect to an arbitrary translation. This can be seen by the fact that a translation by a vector $\mathbf{u}$ in real space multiplies each product of Fourier coefficient in (3) by a phase $\exp(i[\mathbf{k}_1 + \ldots + \mathbf{k}_n]\cdot\mathbf{u})$, which must be equal to unity for the free energy to remain unchanged. An important consequence of this form of Eq. (3), for a generic $F$, is that if $\rho(\mathbf{r})$ is a minimum of *F*, then the set *L* of wave vectors for which $\rho(\mathbf{k}) \neq 0$ is closed under addition, unless Bragg peaks are extinguished due to symmetry. Mermin[24] showed that a generic free energy will be linearly unstable if $\rho(\mathbf{k}) = 0$ at a wave vector **k** that is a linear combination of wave vectors already in *L*, allowing the appearance of a nonzero Bragg peak at **k** to lower the energy. In this sense, the set *L* extends the nature of the reciprocal lattices of periodic crystals to quasiperiodic crystals. We therefore continue calling *L* a *lattice* of wave vectors also in the general case of quasiperiodic crystals. The lattice *L* has the algebraic structure of a finitely generated *Z*-module, and is sometimes called the *Fourier module*. A *Z*-module is like a vector space, only that the scalars are taken from a ring—the integers—rather than a field.

---

[†] By rotations we mean both proper and improper rotations—mirror reflections and the inversion of space—making up the full $d$-dimensional orthogonal group $O(d)$. Naturally, we are mainly concerned with $d = 2$ or 3.

[‡] Some authors require crystals to possess so-called "forbidden symmetries" in order to be regarded as quasicrystals. It is now understood that such a requirement is inappropriate. See Ref. [22] for details, and Ref. [23] for simple examples of square and cubic quasicrystals.

We note in passing that in the immediate years following the discovery of quasicrystals, expansions like Eq. (3), truncated at *n* = 3 or 4, were used to calculate the free energies of different structures in an attempt to explain the stability of quasicrystals. Kalugin, Kitaev, and Levitov [KKL],[25] who extended the famous work of Alexander and McTague[26] explaining why most crystals are *bcc*, even showed that the icosahedral quasicrystal can have a lower free energy than the competing *bcc* phase. But then, Gronlund and Mermin[27] showed that the addition of a quartic term to the cubic free energy of KKL reverses the outcome of the calculation, establishing the *bcc* phase as the favored one. This approach is now being revisited in light of the discovery of quasicrystals in soft matter (for a review see Ref. [28]), where truncated expansions of the free energy might be more valid. It has already yielded an explanation for the thermodynamic stability of a certain class of 2-dimensional soft-matter quasicrystals with dodecagonal symmetry.[29]

Here, rather than calculating the free energies of particular structures, we wish to make use of the free energy expansion (3) to find a general characterization of the order-parameter space. We start by asking what is the relation between two degenerate minimum states of a *generic* free energy.§ In other words, we wish to find the conditions for two different symmetry-broken states to be indistinguishable, as far as the free energy (3) is concerned. It follows directly from the form of Eq. (3) that if two such states—characterized by two different density functions, $\rho(\mathbf{r})$ and $\rho'(\mathbf{r})$—are both minima of *F*, their Fourier coefficients must satisfy the equalities

$$\forall \mathbf{k}_1,\ldots,\mathbf{k}_n \in L: \quad \rho'(\mathbf{k}_1)\cdot\ldots\cdot\rho'(\mathbf{k}_n) = \rho(\mathbf{k}_1)\cdot\ldots\cdot\rho(\mathbf{k}_n), \quad (4)$$

for any *n*, whenever $\mathbf{k}_1 + \ldots + \mathbf{k}_n = 0$. These products are the so-called *structure invariants* that are used as the basis for solving the phase problem in crystallography. They are nothing but the Fourier-space version of the statement that the density autocorrelation functions, of arbitrary order *n*,

$$C^{(n)}(\mathbf{r}_1,\ldots,\mathbf{r}_n) = \lim_{V\to\infty} \frac{1}{V} \int_V \rho(\mathbf{r}_1 - \mathbf{r})\ldots\rho(\mathbf{r}_n - \mathbf{r}) d\mathbf{r}, \quad (5)$$

all give the same values for both states. We used two-point correlations in Eq. (1) to indicate the existence of long-range order. Now we see that for two states to have the same order—both being minima of the same generic free energy—they must agree on all their *n*-point correlations for arbitrary *n*. Following Rokhsar, Wright, and Mermin,[30] we say that two such states are *indistinguishable*. The term *homometric* is used in crystallography to describe two distinct crystals that share the same 2-point correlations, thus producing the same diffraction diagram, yet differ in their higher-order spatial correlations.

The conditions for indistinguishability, stated either in real space (5) or in Fourier space (4), seem quite impractical, as they require one to compare infinitely many correlations. Fortunately, the statement of indistinguishability in Fourier space can be greatly simplified,[29-31] at least in generic situations, as follows. First we note that the densities in real space are real and therefore the Fourier coefficients satisfy $\rho(-\mathbf{k}) = \rho^*(\mathbf{k})$. For 2-point correlations, Eq. (4) is then a statement of the identity of the *magnitudes* of the Fourier coefficients of indistinguishable densities, or equivalently of the fact that two indistinguishable densities must produce identical diffraction diagrams. One can therefore associate a (real) phase $\chi(\mathbf{k})$ with each Bragg peak **k**, such that

$$\forall \mathbf{k} \in L: \quad \rho'(\mathbf{k}) = e^{2\pi i \chi(\mathbf{k})} \rho(\mathbf{k}). \quad (6)$$

Rokhsar, Wright, and Mermin[29,30] called $\chi(\mathbf{k})$ a gauge function, in analogy to the gauge functions in electrodynamics, which can change the phase of a wave function without changing any of its observables. From the fact that the densities are real, together with Eq. (6), it immediately follows that $\chi(-\mathbf{k}) = -\chi(\mathbf{k})$. This, together with the indistinguishability condition (4) for the equality of 3-point correlations, leads to the requirement that $\chi$ be a linear function of the wave vectors in *L*,

$$\forall \mathbf{k}_1, \mathbf{k}_2 \in L: \quad \chi(\mathbf{k}_1 + \mathbf{k}_2) = \chi(\mathbf{k}_1) + \chi(\mathbf{k}_2). \quad (7)$$

The surprising consequence of this last requirement is that for generic quasiperiodic densities it automatically ensures the equality of all higher-order correlations. To see this, simply substitute Eq. (6) into Eq. (4), and make use of the linearity property (7). Thus, we need only check the equality of the 2-point and 3-point correlations in order to determine whether two quasiperiodic densities are indistinguishable, rather than having to examine infinitely many correlations as in Eqs. (4) and (5).** The bookkeeping for this is particularly simple in Fourier space: *Two quasiperiodic densities are both minima of the same generic free energy, termed indistinguishable, if and only if their Fourier coefficients are related by a linear gauge function* $\chi(\mathbf{k})$ *as in Eq.* (6). We should emphasize that this statement, which was recently proven in a more general setting,[33] is true only for generic quasiperiodic densities. For example, Grünbaum and Moore[34] constructed non-generic examples of distinct structures that agree on all their *n*-point correlations up to *n* = 5, but disagree on their 6-point correlations.

It turns out that there is an even deeper surprise hidden behind the italicized statement above, which is revealed by considering the space of all possible gauge functions $\chi(\mathbf{k})$. Because the set of all gauge functions can be used, through Eq. (6), to map a particular minimum free-energy state exactly onto all others, the space of all gauge functions is isomorphic to the space of all minimum free-energy states. Furthermore, because gauge functions are linear, one can uniquely express $\chi(\mathbf{k})$ by specifying its values $\chi_i = \chi(\mathbf{b}^{(i)})$ on a chosen basis for $L$, consisting of $D$ linearly independent wave vectors (over the integers), $\mathbf{b}^{(i)}$ with $i = 1,\ldots,D$. Any set of $D$ real numbers $\chi_i$ specifies a unique gauge function, and any gauge function is uniquely expressed as a set of $D$ numbers (once a particular basis has been chosen). Thus the space of all gauge functions is a $D$-dimensional vector space $V^*$ over the real numbers.[35] Now, here comes the surprise: We have just shown that the symmetry of the free energy is given by gauge functions $\chi(\mathbf{k})$, which are equivalent in an abstract sense to vectors $(\chi_1,\ldots,\chi_D)$ in a $D$-dimensional space. Yet, the symmetry of the liquid phase consists of the set of all $d$-dimensional rigid translations $\mathbf{u} = (u_1,\ldots,u_d)$, where in general $D \geq d$. Thus, *the free energy, in general, when restricted to quasiperiodic functions, has more symmetry than we had initially presumed*. We are still considering a symmetry-breaking phase transition, yet before we break the symmetry, we must step back and realize that the actual symmetry of the free energy consists not only of rigid translations of space. Rigid translations form only part of what one can achieve with gauge functions, using Eq. (6). This situation is possible because we have restricted the definition of gauge functions to a countable set $L$ of wave vectors $\mathbf{k}$ at which $\rho(\mathbf{k}) \neq 0$. For $D > d$, even though this set is *dense*, one cannot extend the gauge functions to functions that are linear for all *continuous* values of $\mathbf{k}$.

---

§ By "generic" we mean roughly that the free energy will not have accidental degeneracies or other peculiar features that could be undone by small variations of its parameters.

** Had we not subtracted the average density in the definition of $\rho(\mathbf{r})$, the condition (4) for *n* = 1 would have required us to verify, in addition, that the two densities have the same average.

One can decompose any given gauge function into a pure $d$-dimensional rigid translation, given by the $d$ components of a translation vector $\mathbf{u}$, and, if $D > d$, a remaining contribution $\varphi(\mathbf{k})$, called a *phason*, that affects only the *relative phases* of the Fourier coefficients, leaving some chosen origin fixed. This decomposition is achieved through a change of basis in the space $V^*$ of gauge functions,

$$\chi(\mathbf{k}) = \sum n_i \chi_i = \frac{\mathbf{u} \cdot \mathbf{k}}{2\pi} + \varphi(\mathbf{k}), \text{ where } \mathbf{k} = \sum n_i \mathbf{b}^{(i)}. \quad (8)$$

For periodic crystals[31] $\varphi(\mathbf{k}) = 0$, and the translation vector is simply given by $\mathbf{u} = \sum \chi_j \mathbf{a}^{(j)}$, where the vectors $\mathbf{a}^{(j)}$ are the usual real-space dual vectors, satisfying $\mathbf{a}^{(j)} \cdot \mathbf{b}^{(i)} = 2\pi \delta_{ij}$. We demonstrate the decomposition of a gauge function into a translation and a phason below, for an operation that leaves the ordered state invariant.

### 3.2.3. What remains of the broken translational symmetry?

The key to understanding what remains of the full symmetry of the free energy, when that symmetry is broken, is the fact that the gauge function appears as a multiplicative phase in Eq. (6). Thus, transforming a density $\rho$ with a gauge function $\chi$, as in (6), leaves it invariant if and only if $\chi$ is integral-valued on all the wave vectors (the factor of $2\pi$ having been conveniently taken out by the definition). If the crystal is periodic, then all the possible integral-valued gauge functions correspond, through the relation (8), exactly to the set of translations leaving the crystal invariant. In that case, the full symmetry group $G$, which is the set of all rigid translations of $d$-dimensional space, is broken into a discrete and periodic lattice of translations. For quasicrystals, the effect of transforming $\rho$ with an integral-valued gauge function $\chi$ is more elaborate, combining translations with phasons. The latter appear in the form of nontrivial spatially correlated rearrangements of tiles (in tiling models) or of the atoms (in real crystals). Such a symmetry operation is demonstrated in Fig. 2, using the Penrose tiling.

### 3.2.4. What is the order-parameter space?

Let us follow Dräger and Mermin[35] and slightly formalize the observations of the previous subsections. Gauge functions are useful in describing the relations between the different symmetry-broken minimum free-energy states. Gauge functions form a vector space $V^*$ of all real-valued linear functions on the lattice $L$, and because $L$ has rank $D$, $V^*$ is a $D$-dimensional vector space over the real numbers. The space $V^*$ encodes the full symmetry of the free energy $F$. The space $V^*$ contains, as a subspace, all the integral-valued linear functions on $L$. We denote this subset—which has the algebraic structure of a rank-$D$ Z-module, just like $L$ itself—by $L^*$. Gauge functions in $L^*$ leave the minimum free-energy state invariant, thus encoding the symmetry that remains in the ordered phase. Gauge functions that belong to the quotient space $V^*/L^*$ transform the ordered state described by $\rho$ into a different, yet indistinguishable, ordered state described by some other density function $\rho'$. Thus, $V^*/L^*$—the space of all linear functions on $L$ with real values modulo the integers—is the order-parameter space. One can parameterize all the degenerate ordered states by a set of $D$ numbers $0 \leq \chi_i < 1$, $i = 1,\ldots,D$. Geometrically, this can be viewed as a simple $D$-torus, or equivalently as a $D$-dimensional unit cube with periodic boundary conditions. Arithmetic, involving gauge functions, is performed from here on modulo the integers, to remain within the order-parameter space $V^*/L^*$. We denote equality to within an additive integer by the symbol ' $\equiv$ '.

### 3.2.5. What remains of the broken rotational symmetry?

We are now in a position to describe the rotational symmetry of the ordered phase. But first, we should ask ourselves what we mean when we say that a crystal has a certain rotational symmetry. After all, we have redefined the notion of a crystal. Is it possible to avoid a redefinition of the notion of symmetry?

In the case of a periodic crystal we mean that the rotation leaves the density of the crystal invariant to within a translation. This means that after rotating we may need to apply a translation, after which the points of the rotated crystal will exactly coincide with the points of the original crystal. The densities of quasicrystals, however, in general possess no such symmetries. In fact, it is a nice exercise to show that if a 2-dimensional crystal contains more than a single point, about which an $n$-fold rotation ($n > 2$) brings it into perfect coincidence with itself, the crystal is necessarily periodic. So a crystal with, say, 5-fold symmetry cannot contain more than a single point of "exact" 5-fold symmetry, and if we were given a finite piece of that crystal there is a good chance that we would never find that point. We can therefore no longer rely on the criterion of invariance to within a translation as a definition of crystal symmetry.

Our discussion above, on the order-parameter space, holds the key to resolving this problem, and thereby adapting the notion of symmetry to the age of quasicrystals. For periodic crystals, the property of invariance to within a translation simply means that the rotation maps a particular ordered state into one of the other states in the order-parameter space. This can be generalized to any quasiperiodic crystal by saying that *a symmetry operation leaves the crystal indistinguishable rather than invariant to within a translation*. Thus, a symmetry operation need not leave the density $\rho(\mathbf{r})$ invariant, but rather all of its autocorrelations, as expressed in Eqs. (4) or (5), must be left invariant.

Indeed, certain rotations, when applied to a quasiperiodic crystal, take it into one that looks very much like the original unrotated crystal, similar to what we saw in Fig. 2 after applying an integral-valued gauge function. This is demonstrated in Fig. 3, where a blue Penrose tiling is placed over an identical red one, and is then rotated. Nothing interesting happens until a 10-fold (36 degree) rotation is completed and a fair fraction of the vertices of the two tilings coincide (top figure). The blue tiling is then translated by some amount until whole regions, of about 10 to 20 tiles across, coincide (middle figure). Finally, the blue tiling is translated even further until regions of the order of the whole observed patch coincide. Between the coinciding regions there always remain strips, containing tiles that do not match, as we saw earlier in Fig. 2. If we could see the entire infinite tiling we would observe that any bounded region in the rotated tiling can be found in the unrotated tiling, but the larger the region the further away one has to look in order to find it. Even so, there is no translation that will bring the whole infinite blue tiling into full coincidence with the infinite red one, as the Penrose tiling contains not even a single point of "exact" 10-fold symmetry. One will always need to add a phason component, rearranging the tiles in a correlated manner throughout the whole tiling, in order to obtain full coincidence.

The collection of all rotations and reflections $g \in O(d)$ that are symmetries of a given crystal form the *point group* of the crystal. To fully specify the symmetry of the crystal—given by its *space group*—we need to indicate exactly in what way the original and rotated versions of the crystal differ. If the crystal is periodic, where indistinguishability reduces to invariance to within a translation, all we need is to specify exactly what translation $\mathbf{t}_g$ needs to follow each rotation to leave the crystal invariant. If the crystal is quasiperiodic, simple translations, in general, are insufficient and the required information is encoded in a special gauge function, $\phi_g(\mathbf{k})$, known as a *phase function*,[29,30] relating $\rho(g\mathbf{r})$ and $\rho(\mathbf{r})$ through their Fourier coefficients,

$$\forall \mathbf{k} \in L: \quad \rho(g\mathbf{k}) = e^{2\pi i \phi_g(\mathbf{k})} \rho(\mathbf{k}). \quad (9)$$

This so-called *point-group condition* (9) is used to decide whether a rotation $g$ is in the point group of the crystal. It also imposes immediate constraints between the phase functions associated with two point-group operations $g$ and $h$ and their product $gh$, known as the *group-compatibility conditions*,

$$\forall g, h: \quad \phi_{gh}(\mathbf{k}) \equiv \phi_g(h\mathbf{k}) + \phi_h(\mathbf{k}). \quad (10)$$

The possible solutions to these coupled equations (10) are used to find all distinct sets of phase functions for a given point group, using various equivalence criteria. This provides a very simple and elegant approach for calculating, and then listing, all the possible space group types.[31,32] This approach has been generalized for dealing with magnetic symmetry[36] and color symmetry[37] in quasicrystals, and also for treating modulated crystals and composite structures[††].[39] For an introduction see Ref. [40]. We should remind the reader that the idea of describing the symmetry of crystals in Fourier space was suggested long ago by Bienenstock and Ewald[41], but its usefulness became clear only after the discovery of quasicrystals. We also note that expressions like the group-compatibility conditions (10) are known from cohomology theory, where they are called *cocycles*, thus making another surprising connection between crystals, diffraction theory, and abstract mathematics.[42]

Finally, the point-group condition (9) allows one directly to predict the existence of extinctions in the diffraction diagram of a crystal.[43] These are wave vectors in the lattice $L$ that are expected to carry Bragg peaks, owing to the closure of $L$ under addition. Nevertheless, they are extinguished due to symmetry. Indeed, if a wave vector $\mathbf{k}$ is invariant under an operation $g$ ($g\mathbf{k} = \mathbf{k}$), then according to Eq. (9) $\rho(\mathbf{k})$ must vanish unless $\phi_g(\mathbf{k}) \equiv 0$. This is essentially the whole argument, except for the fact that one can show that for a given space group type, the value of $\phi_g(\mathbf{k})$ at such wave vectors $\mathbf{k}$ that are invariant under $g$, is independent of the specific ordered state that is chosen within the order-parameter space.

## 4. What are the elementary excitations?

We shall not give here a complete account of the elementary excitations in quasiperiodic crystals, as many discussions and reviews can be found elsewhere.[44-46] We only wish to describe how they emerge naturally in our discussion of symmetry breaking.[9] As a direct consequence of the breaking of continuous symmetry, known as the *Goldstone theorem*, one can introduce long-wavelength deformations of the ordered state that cost very little energy. In the infinite-wavelength limit these deformations gradually reduce to a pure gauge transformation, having no energy cost at all. In the case of periodic crystals, these *Goldstone modes* are the well-known phonon modes, or sound-wave excitations. At any given wave vector $\mathbf{q}$, there will be $d$ phonon modes, because there are $d$ distinct polarizations, or $d$ components in the translation vector $\mathbf{u}$. More generally, in a quasiperiodic crystal, we expect to have $D$ polarizations at any wave vector $\mathbf{q}$, coming from the richer possibilities afforded by the gauge functions. As in Eq. (8), an appropriate basis can always be chosen that reflects the fact the first $d$ modes are the usual *phonon modes*, arising from the pure translational components of the gauge function, while the remaining $D - d$ modes are called *phason modes*.

We have just seen that gauge functions $\chi(\mathbf{k})$ transform one ordered state into another at no energy cost. After choosing a basis, the gauge functions are specified by a set of $D$ numbers $(\chi_1, \ldots, \chi_D)$. To introduce a deformation in the otherwise perfectly ordered state, we allow the gauge function to vary slowly in space. Thus, locally at each point we have one particular ordered state, but as we move from one region in the crystal to another, we gradually explore the order-parameter space. To formalize this we introduce the *order-parameter field* (not to be confused with the order parameter discussed in Sec. 2), consisting of $D$ components $\chi_i(\mathbf{r})$. If the order-parameter field is constant in space it reduces to a simple gauge function and no energy is required. If it varies slowly in space, say as $\exp(i\mathbf{q}\cdot\mathbf{r})$, there will be an energy cost coming from the gradients, whose leading contribution is proportional to $|\nabla\chi_i(\mathbf{r})|^2 \sim |\mathbf{q}|^2$, continuously approaching zero at the infinite wave-length (or zero wave-vector) limit.

---

[††] These quasiperiodic crystals were known before the discovery of quasicrystals, but did not pose any threat to the periodicity paradigm, as they are formed by slightly modifying an underlying periodic crystal. Their crystallography was treated by de Wolff, Janner, and Janssen[38] by embedding them in spaces of $D$ dimensions, where periodicity is recovered.

It is common, although not always necessary, to describe the order-parameter field in the phonon-phason basis, as in Eq. (8). In that case, the phonon distortion is expressed using the usual $d$-dimensional displacement field $\mathbf{u}(\mathbf{r})$, describing the small deviations of the density from its equilibrium position at $\mathbf{r}$. In many simple situations it just so happens that $D = 2d$. For icosahedral crystals in 3 dimensions the rank is 6, and in 5-, 8-, 10-, and 12-fold crystals in 2 dimensions the rank is 4. In these cases the number of phason components is also equal to the physical dimension and one often describes the phason distortion using a second $d$-dimensional field $\mathbf{w}(\mathbf{r})$. The full $D$-component order-parameter field, can then be expressed as

$$\chi(\mathbf{r};\mathbf{k}) = \sum n_i\chi_i(\mathbf{r}) = \frac{1}{2\pi}\mathbf{u}(\mathbf{r})\cdot\mathbf{k} + \frac{1}{2\pi}\mathbf{w}(\mathbf{r})\cdot\tilde{\mathbf{k}}, \quad (11)$$

where $\tilde{\mathbf{k}}$ is a $d$-dimensional vector, obtained from the original vector $\mathbf{k}$ by a linear transformation that permutes its coefficients $n_i$, so that the dot product with the phason field $\mathbf{w}(\mathbf{r})$ induces a relative phason distortion, rather than a simple displacement. For example, for 5-fold and 10-fold quasicrystals in 2-dimensions, if $\mathbf{k}$ is given by the four integers $(n_1,n_2,n_3,n_4)$, in the standard basis of four wave vectors $\mathbf{b}^{(i)}$ separated by 72 degrees, then $\tilde{\mathbf{k}}$ is given by $(n_2,n_4,n_1,n_3)$ in the same basis.[46] It should be noted that the phonon-phason basis, in general, does not diagonalize the free energy, in the sense that even in the harmonic approximation there is a coupling between the two fields.

Much like phonons, phasons are low-energy excitations of the quasicrystal, affecting its hydrodynamic description and elastic properties, only that instead of encoding the fluctuations of atoms away from their equilibrium positions, they encode relative changes in the positions of atoms, called *phason flips*. In Figs. 2 and 3 we saw long strips of correlated phason flips. In a real quasicrystal, thermal fluctuations may induce uncorrelated phason flips, and upon lowering the temperature there often remains frozen-in disorder in the form of uncorrelated phason flips. This leads to a particular broadening of the Bragg peaks and the addition of diffuse scattering. Indeed, these effects of the phason excitations are detected regularly in experiments,[45] including even the direct observation of individual phason flips using high-resolution TEM,[47] or by studying large-scale metamaterials such as phonotic quasicrystals.[48] Nevertheless, phasons are a continuing source of interesting and surprising puzzles, and ongoing debate.[46]

## 5. What are the topological defects?

As in the previous section, we shall not give here a full account of the topological defects in quasiperiodic crystals,[49] and again only show how they enter into the framework of the liquid-to-crystal symmetry breaking transition. Equipped with our convenient parameterization of the order-parameter space as a $D$-torus, we can proceed to understanding and classifying the different topological defects, or *dislocations*, that can form in the otherwise perfectly ordered state. Again, we use the same $D$-component order-parameter field, $\chi_i(\mathbf{r})$, and gradually explore the order-parameter space as a function of position, but this time going in a loop around the core of the dislocation rather than going in a straight line as before. To do so, it is convenient to express the position vector $\mathbf{r}$ in polar coordinates $(r,\theta)$ and use a gauge function that depends only on the angle coordinate $\chi_i(\theta)$.

The topological nature of the defect[50] is related to the fact that it cannot be made to disappear by local structural changes. For this to be the case, as we traverse in a loop around the position of the dislocation, say the origin, we follow a curve in the order-parameter space that winds around the $i^{th}$ direction of the $D$-torus $n_i$ times, and returns back to the original point in order-parameter space. This yields a crystal that is everywhere only-slightly distorted from the ordered state, except near the core of the dislocation. It is accomplished by taking $\chi_i(\theta) = n_i\theta$. Thus, the most general dislocation is characterized by a set of $D$ integers $(n_1,\ldots,n_D)$, which for a periodic crystal reduces to the familiar $d$-dimensional Burgers vector. Not surprisingly, the set of all dislocations forms a rank-$D$ Z-module—the so-called *homotopy group* of the $D$-torus.[50] A pair of dislocations—introduced in this manner into the dodecagonal ordered state of a certain free energy, introduced by Lifshitz and Petrich[51]—are shown in Fig. 4 after a short relaxation time.

Once we understand how the order-parameter field $\chi_i(\mathbf{r})$ encodes the information about the possible existence of dislocations, we can readily extract this information for any given structure. Figure 4 demonstrates how one can actually do this. The procedure shown may be carried out in "real-time", while simulating the dynamics of a quasicrystal, so that dislocation motion can be followed quantitatively as the system evolves. Barak and Lifshitz[52] used this technique to study such questions as the climb velocity of dislocations under strain, the pinning of dislocations by the underlying quasiperiodic structure under conditions of weak diffusion, and the relaxation of phason strain as two dislocations of opposite topological sign, like the ones in Fig. 4, merge and annihilate each other. They did so by using particular dynamics, based on a simple relaxation of the Lifshitz and Petrich[51] free energy. Freedman *et al.*[48] used a similar procedure to analyze experimental images of dislocations in photonic quasicrystals.

## 6. Summary and outlook

I have presented an overview of quasicrystals, within the framework of a symmetry-breaking phase transition, which may not yet be complete, but where much is now known and well understood. I avoided the use of $D$-dimensional spaces, where $D > d$ vectors can be made mutually orthogonal, thus allowing one to rely on all the known concepts of periodic crystals, yet in an abstract high-dimensional space.[38] Instead, I followed an approach, formulated in $d$-dimensional physical space—originally introduced by Rokhsar, Wright, and Mermin,[30]—insisting on forming a new understanding of old fundamental notions, and adapting them to the age of quasicrystals. I have shown that Bragg peaks and their associated gauge functions are all that is required to address the four questions, introduced at the outset. The gauge functions are used to describe the broken symmetry in quasi-periodic crystals, to characterize the order-parameter space, and to construct an order-parameter field for studying defects and excitations. They contain all the necessary information, in their $D$ components (once a particular basis is chosen), and offer an attractive alternative to the need for embedding the crystal itself in a $D$-dimensional space.

Although much progress has been achieved so far in the study of quasicrystals, it is clear that there is still even more to be done. I therefore believe that there are many years of exciting research still ahead before crystallography becomes, once again, a mature science awaiting its next revolution.

## Acknowledgements

I am grateful to my teachers at Cornell University, Neil Ashcroft, Veit Elser, Chris Henley, Jim Sethna, and first and foremost, my adviser David Mermin, for instilling in me the basis for much of what I have described here. I thank Mike Cross and David Mermin for their thoughtful remarks on an earlier draft of this manuscript. This work is supported by the Israel Science Foundation through Grant No. 556/10.

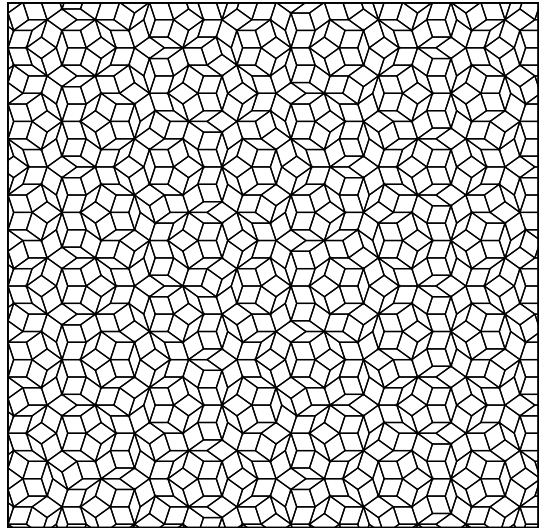

**Figure 1**. A small section of the rhombic Penrose tiling,[12] containing two kinds of rhombic tiles—with a small angle of 36 degrees and of 72 degrees. These two types of tiles are arranged in a very specific and ordered manner to produce a tiling that is quasiperiodic. If we were to place atoms at, say, the vertices of the tiling and perform a diffraction experiment, we would see a 10-fold symmetric Bragg peak diffraction diagram, as was originally done by Mackay.[14]

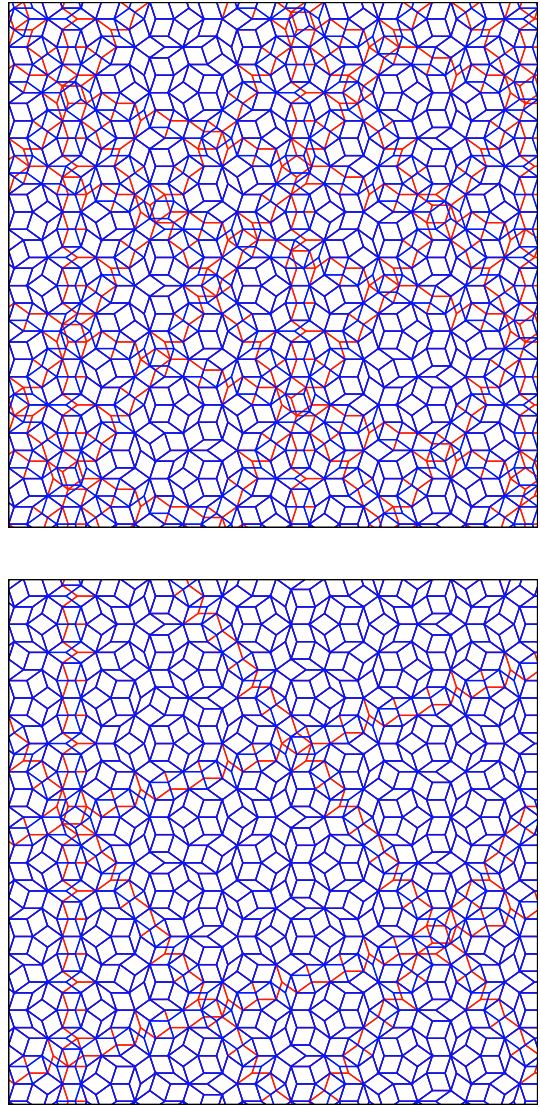

**Figure 2.** (Color) The Penrose tiling of Fig. 1 in red, covered by an exact copy of itself in blue shifted by a small amount (top), and shifted by a larger amount (bottom). Red tile edges are visible in regions of mismatch. An increase in the coincidence of vertices and tiles is clearly visible the greater the translation. Viewing the images of superimposed tilings with a slight defocusing enhances the appearance of highly correlated lines containing all the mismatched tiles, which must be rearranged to obtain the original tiling. These rearrangements are effected by the phason components of the integral-valued gauge function.

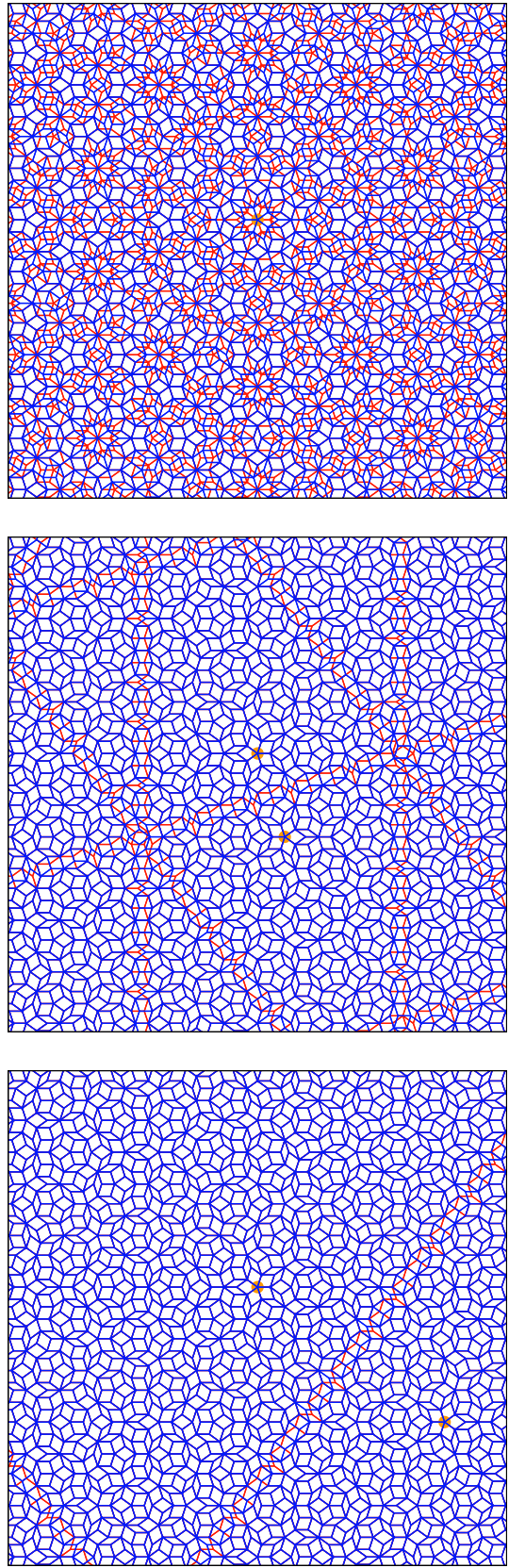

**Figure 3.** (Color) The Penrose tiling of Fig. 1 in red, covered by an exact copy of itself in blue rotated by 36 degrees (top); then shifted by a small amount (middle); and finally shifted by a larger amount (bottom). Red tile edges are visible in regions of mismatch. As in Fig. 2, an increase in the coincidence of vertices and tiles is clearly visible the greater the translation, following the 10-fold rotation. Even so, there is no translation that will bring the whole infinite red tiling into full coincidence with the infinite blue one, as the Penrose tiling contains not even a single point of "exact" 10-fold symmetry. The gauge functions relating the red and blue tiles in each frame differ from each other by an integer-valued gauge function from $L^*$. Orange dots mark the original point of rotation on the two tilings.

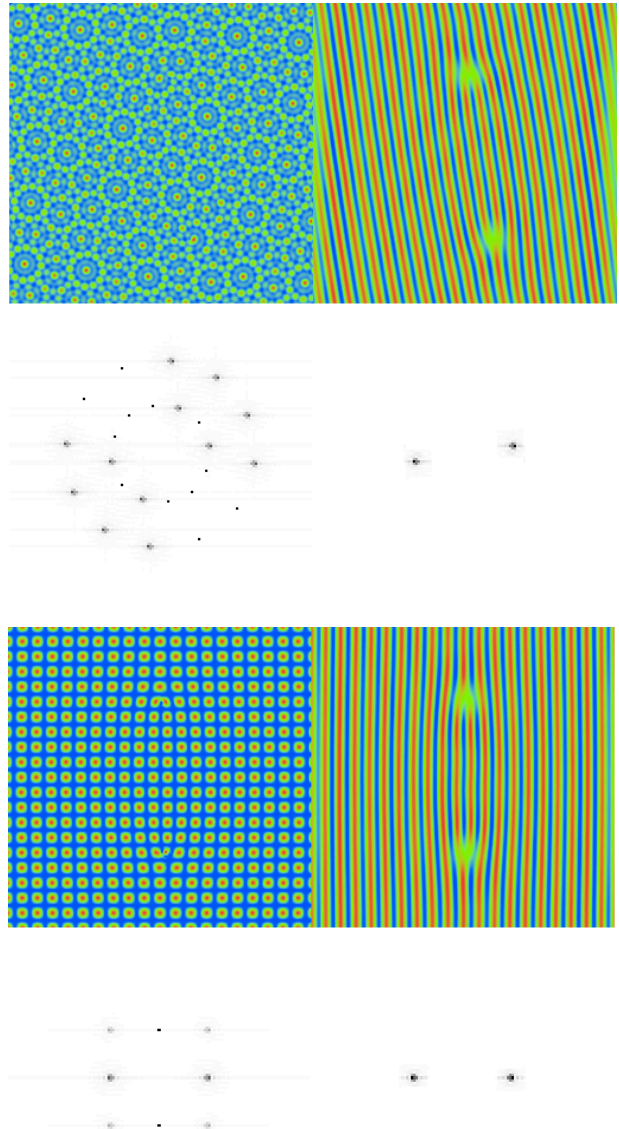

**Figure 4.** (Color online) Top-left (in both images): Snapshot of the numerical solution of the Lifshitz-Petrich equation[51] showing a dodecagonal and a square pattern, a short time after a pair of dislocations had been injected at two separate positions, with Burgers vectors (1,0,0,0), (-1,0,0,0) and (1,0), (-1,0), respectively. Bottom-left: Fourier transform of the pattern. Note the fuzzy Fourier coefficients, containing the information about the angle-dependent local gauge transformation. Bottom-right: A pair of fuzzy Bragg peaks, filtered out of the full Fourier image. Top-right: Inverse Fourier transform of the filtered peaks revealing the dislocation pairs.